\documentstyle[12pt]{article}

\newcommand{\bmat}{\left(\begin{array}}
\newcommand{\emat}{\end{array}\right)}
\def\NPB#1#2#3{Nucl. Phys. B{#1} (19#2) #3}
\def\PLB#1#2#3{Phys. Lett. B{#1} (19#2) #3}

\def\PRD#1#2#3{Phys. Rev. D{#1} (19#2) #3}

\def\yzero{\smash{\hbox{$y\kern-4pt\raise1pt\hbox{${}^\circ$}$}}}

\def\-{\hphantom{-}}
\def\s2{\frac{1}{\sqrt2}}

\def\beq{\begin{equation}}
\def\eeq{\end{equation}}
\def\beqa{\begin{eqnarray}}
\def\eeqa{\end{eqnarray}}

\def\IF{\relax{\rm I\kern-.18em F}}
\def\II{\relax{\rm I\kern-.18em I}}
\def\IP{\relax{\rm I\kern-.18em P}}
\def\IC{\relax\hbox{\kern.25em$\inbar\kern-.3em{\rm C}$}}
\def\IR{\relax{\rm I\kern-.18em R}}

\def\Dsl{\,\raise.15ex\hbox{/}\mkern-13.5mu D} 
\def\IZ{Z\kern-.4em  Z}

\topmargin
-1.5cm
\textwidth
15.5cm
\textheight
24cm
\oddsidemargin
0.7cm
\evensidemargin
1.2cm

\begin{document}

\makeatletter
\@addtoreset{equation}{section}
\makeatother
\renewcommand{\theequation}{\thesection.\arabic{equation}}
\pagestyle{empty}
\rightline{\tt hep-ph/9804250}
\vspace{0.5cm}
\begin{center}
\LARGE{Non-Perturbative effects from orbifold constructions \\[10mm]
\footnote{To appear in the proceedings of the
Workshop on Phenomenological Applications of String Theory (PAST97),
ICTP, Trieste (Italy) , October 1997.}}
\large{G.~Aldazabal}
\small{
 CNEA, Centro At\'omico Bariloche,\\[-0.3em]
8400 S.C. de Bariloche, and CONICET, Argentina.\\[1mm] }
\small{\bf Abstract} \\[7mm]
\end{center}

\begin{center}
\begin{minipage}[h]{14.0cm}
We indicate how consistent heterotic orbifold compactifications,
including non perturbative information, can be constructed.
We first analyse the situation in six dimensions,   $N=1$, where strong
coupling effects, implying the presence of five branes, are better known.
We show that anomaly free models can be obtained even the usual
modular invariance constraints are not satisfied.  The perturbative massless
sector can be computed explicitly from the  perturbative mass formula
subject  to an extra shift in the vacuum energy.  Explicit examples in
$D=4$, $N=1$ are presented.  Generically, examples  exhibit non
perturbative transitions leading to gauge  enhancement and/or where the
number of chiral generations changes.
\end{minipage}
\end{center}
\newpage
\setcounter{page}{1}
\pagestyle{plain}
\section{Introduction}
The main aim of this  talk, based on work presented in Ref.[1],
 is to indicate  how consistent models, including non perturbative effects,
can be constructed in the framework of heterotic string
orbifolds.  Interestingly enough, some of these models exhibit features
like gauge symmetry enhancing or transitions changing the number of
fermionic generations.  Such kinds of phenomena are not possible in
perturbation theory and imply a drastic change in our way of approaching
string inspired phenomenology.

Due to their simplicity, perturbative heterotic orbifolds
\cite{dhvw,orbi} have proven  to be a very powerful tool in building
"semirealistic stringy inspired", effective low energy models (Standard
model like, higher level StrinGuts etc.).  Not only the gauge group and
matter multiplets may be easily obtained (and somewhat controlled) but also
the structure of Yukawa couplings, symmetry breaking patterns can be
easily studied etc..
The full partition function may be constructed in these
cases.

We will not be able to go that far in the non perturbative case and we will
concentrate on the structure of the massless spectrum.

Our first scope will be to incorporate non perturbative contributions
in six dimensional orbifold compactifications of the heterotic,
both $E_8\times E_8$ and $SO(32)$,  string theories.
  If non perturbative phenomena
are to be included, this seems a good road  before
descending to the more involved and less known four dimensional case. In
fact, we will show how relevant information in four dimensions may be
obtained from six.

$D=6$ is undoubtedly interesting in itself. $N=1$ theories are chiral
and consistency, that is, anomaly cancellation, constraints
the allowed theories severely. Moreover, even if still incomplete,
many non perturbative effects are quite  well understood in six
dimensions.  In particular, examples have been derived from different
approaches as Type IIB orientifolds F-theory and M-theory.

 In an  orbifold
compactification $Z_M$  symmetry is divided out.
Acting on the (complex) bosonic transverse coordinates,
the $Z_M$ twist generator $\theta$ has eigenvalues $e^{2 \pi i\, v_a}$. In
$D=6$ $v_a$ are the components of $v=(0,0,\frac1M,-\frac1M)$ and   $M$
can take the values $M=2,3,4,6$.
The embedding of $\theta$ on the gauge
degrees of freedom is usually performed by a shift $V$, such that
$MV$ belongs to the $E_8\times E_8$ lattice $\Gamma_8 \times \Gamma_8$ or to
the $Spin(32)/Z_2$ lattice $\Gamma_{16}$.

In perturbative string theory, this shift is restricted by the
modular invariance constraint
\beq
 M\, (V^2-v^2)={\rm even}
\label{dos}
\eeq

The spectrum for each model is subdivided in sectors. There are $M$ sectors
twisted by $\theta^j, j=0,1,\cdots , M-1$. Each particle state is created by
a product of left and right vertex operators  $L\otimes R$. At a generic point
in the four-torus moduli space, the massless states follow from
\beqa
\label{uno}
m_R^2& = & N_R+\frac{1}{2}\, (r+j\,v)^2
+E_j-\frac{1}{2} \quad ;\quad  \nonumber \\
m_L^2& = &  N_L+\frac{1}{2}\, (P+j\,V)^2+E_j-1
\eeqa
Here $r$ is an $SO(8)$ weight with $\sum_{i=1}^4 r_i=\rm odd$
and $P$ is a  gauge lattice vector with $\sum_{I=1}^{16} P^I= \rm even$.
$E_j$ is the twisted oscillator contribution to the zero point energy and it
is given by $E_j=j(M-j)/M^2$. The multiplicity of states satisfying
Eq. ~(\ref{uno}) in a $\theta^j$ sector is given by the appropriate
generalized GSO projections \cite{walton,afiq}.  The gravity multiplet,
a tensor multiplet, charged hypermultiplets and 2 neutral hypermultiplets
(4 in the case of $Z_2$) appear in the untwisted sector. Twisted sectors
contain only charged hypermultiplets.  The generalized GSO
projections are particularly simple in the $Z_2$ and $Z_3$ case since
all massless states survive with the same multiplicity.

Let us come back to equation ~(\ref{dos}). This constraint ensures level
matching. It corresponds to an orbifold version of the global consistency
of the theory ensuring anomaly cancellation. This consistency may be
understood as the vanishing of the total magnetic charge associated to
the antisymmetric tensor field. Namely

$\int  _{X} dH\  =\int  _{X} (F ^2-R^2)=\ 0$
where $H$ is the three-form heterotic field strength with
$dH=tr F^2- tr R^2$
and  $X$ is the compact space.
 For an orbifold $X= T^4 /{Z_M}$ and, since  curvature is
localized at ${Z_M}$ the fixed points, we can restate this equation as
\begin{equation}
Q_{TOT}= \sum _f Q_f=0
\label{qmi}
\end{equation}
where integrals are taken around fixed points $f$.
Equivalently, since
the total Euler number of $X$ is $\int  _{X} R^2=24 $ we have
\begin{equation}
I_{TOT}= \sum _f I_f =24
\label{i24}
\end{equation}
where $ I_f$ is the instanton number at the fixed point.

The issue we want to stress here is that modular invariance requirement
as stated in ~(\ref{dos}) and anomaly cancellation are equivalent. They are
satisfied if there are 24 instantons at fixed points or equivalently if the
magnetic charges at fixed points add up to zero.
 More explicitly, in Ref. [1] it is shown
(for $SO(32)$ case) that
$I_f=l+M E_{\Theta }$
Here
$E_{\Theta }= \sum _{I=1} ^{16} { \frac {1}2 V_I (V_I-1)}$,
$E_{\theta}= \frac {(M-1)}M^2$,  $l$ is an
integer and $V_I$ are the components of the shift $V$. Also, by computing the
curvature at the fixed orbifold point it is found that
\begin{equation}
Q_f=l'+M(E_{\Theta }-E_{\theta })
\label{qf}
\end{equation}
with $l'=l-M-1$. Thus, for $Q_f=0$, Eq. ~(\ref{dos}) is
obtained.

  $Z_M$ orbifolds corresponding to all possible embeddings allowed by
equation ~(\ref{dos}) can be constructed. Indeed, their corresponding
massless
 spectra may be reproduced by application of Index theorems on orbifold (ALE)
 singularities \cite{afiuv} with $I_{TOT}=24$ instantons.

 The question to address now is : Could we still have
 a consistent theory if $I_{TOT} <24$ is allowed, i.e. when $n_B=24-I_{TOT}$
instantons become small?
Since  the dilaton is known to diverge \cite{callan} in such a situation,
non perturbative information is required to answer this question.
In fact, small instantons in both $SO(32)$ or $E_8 \times E_8$ have been
 studied \cite{witsm,dmw} and may be identified as five branes, i.e.,
extended objects with their world volume filling six dimensional space time.
They correspond to type I $D5-branes$ in the $SO(32)$ case and to $M-theory$
five branes for $E_8 \times E_8$.
Five branes act as magnetic sources for the antisymmetric tensor field and
therefore Eq.~(\ref{qmi}) must now read
\begin{equation}
Q_{TOT}= \sum _f Q_f+ n_B=0
\label{qn}
\end{equation}

 We see that, from our discussion about Eqns. ~(\ref{qn}, \ref{i24}), when
five branes are present  the ``modular invariance" constraint on the shift
$V$ must be abandoned.
 This is certainly troublesome, in perturbation theory,  since only
shifts complying with  this constraint ensure anomaly
cancellation.  Other $V$'s would lead to anomalous spectra.  On the other
hand this  should not be surprising when dealing with strong coupling
effects, since modular invariance is a perturbative concept (associated to
the expansion in terms Riemman surfaces spanned in string propagation).

Generically (we will be more precise about this), these five branes are
expected to carry vector, hyper and tensor massless (six dimensional)
multiplets on their world volumes and therefore are expected to contribute
to the total, gauge and gravitational anomaly, of the spectrum.

All these elements suggest a possible positive answer to the above
question.  Perturbative contributions associated to "fat" $I$ instantons and
to $n_B$ five branes, with  $I+ n_B= 24$ , would  contribute to the massless
spectrum such that the whole anomaly could  cancel. In fact, we will see
that this appears to be the case for the situations  where non perturbative
information at hand.

Let us first discuss the perturbative contribution to the massless spectrum.
 This spectrum corresponds to a number
$I_{TOT}<24$ of large instantons. As indicated, the instanton number
is a function of the shift $V$ in the gauge lattice. This $V$ have to
comply wit  a new constraint depending on the number of five branes since
$\sum I_{f}(V)+n_B=24$. For instance, assume that we have the same charge at
each fixed point (this is expected  for $Z_3$ orbifold where all points
are equivalent). Equation ~(\ref{qn}) tells us that $Q_f= -\frac {n_B}n_f$
where $n_f$ is the number of fixed points. Following the steps
that lead us to ~(\ref{qf}) we now obtain
\beq
 M\,
 (V^2-v^2)+2ME_B(f)={\rm even}
\label{lmb}
\eeq
 where we have defined, for
further convenience, $E_B(f)= -M\frac {n_B}{2n_f}$.  This offers us the
result we expected. Moreover, recalling that ~(\ref{dos}) results by
imposing level matching, our result suggests that masses of states could be
obtained as in ordinary perturbative orbifolds by just  modifying the mass
of the left sector states to be
\begin{equation}
m_L^2  =   N_L+\frac{1}{2}\, (P+j\,V)^2+E_j++E_B(j)-1
\label{mlb}.
\end{equation}
In fact, $m_L^2=m_R^2$ leads to ~(\ref{lmb}) with $f$ a fixed point in twisted
sector (j).  We will propose this expression for computing the massless
states in the perturbative sector of general orbifold models containing five
branes. $E_B$ is interpreted as shift in the vacuum energy due to the flux
of the antisymmetric field.  Since in general there will be non equivalent
fixed points we do not expect in general a simple relationship as above
between this energy shift and $n_B$.
 The untwisted sector is obtained by projecting onto invariant states
as usual.

In order to illustrate how this proposal works,  let us consider
the case of smooth $Z_3$ compactifications. This is the simplest case. There
is just one $\theta$ twisted sector with an energy shift to be considered
and nine equivalent fixed points.  Smooth compactification  means that
oscillator modes, needed to blow-up orbifolds singularities  should be
present, thus  $N_L=1/3 $ in ~(\ref{mlb}). For these modes (two at each fixed
point) to be massless it is required that
\beq
\label{lengths}
V^2\ =\   \frac89 \ -\  2\ E_B
\eeq
Thus the maximum shift in the vacuum energy will correspond to
$E_B=\frac49$  (obtained for  $V=0$).  The other extreme case is
$V^2=\frac89$,
in which we have $E_B=0$   corresponding  to  some   modular
invariant (perturbative) models.

Let us consider first the $SO(32)$ heterotic string with  the
class of shifts  $V$  with  $3V\in \Gamma _{16} $  of the form
\beq
\label{shifts}
V\ =\ \frac13 (1,\cdots , 1,0, \cdots ,0)
\eeq
and $m\leq 8$.
The unbroken group is $U(m)\times SO(32-2m)$ and the
untwisted sector contains hypermultiplets transforming as
$({\bf  m}, {\bf
32-2m})$ $+({\bf   \frac{m(m-1)}2}, {\bf 1})$ $+2({\bf 1},{\bf 1})$.  The
twisted sectors need an extra vacuum shift $E_B= \frac{(8-m)}{18}$ and
the mass formula  provides massless hypermultiplets in each twisted sector
transforming  as
\beq
\label{tspec}
({\bf   \frac{m(m-1)}2}, {\bf 1}) \ +\  2({\bf 1},{\bf 1})
\eeq
for $m=0,2,4,6,8$.

There are two other $Z_3$ models with
singlet  moduli in the twisted sector.  One of them, with
shift $V=(\frac23,0, \cdots,0)$,  has gauge group
 $SO(30)\times U(1)$.
The other model has shift $V=\frac16(1, \cdots ,1)$,  $3V$ being a spinorial
weight. The gauge group is $U(16)$. It is thus a $SO(32)$ embedding
without vector structure,
a $Z_3$ analogue to the $Z_2$  orientifolds constructed  in \cite{bso,gp}.

Except for the $m=8$ case, remaining models, as they
stand, have gauge and gravitational anomalies  and the  corresponding shifts
do not fulfill the perturbative modular invariance constraints.  However, it
turns out  that the addition of an appropriate number of five-branes renders
them consistent.  Indeed, one can check that adding $3(8-m)$  five-branes to
the vacua in Eq.~(\ref{shifts}) (12 five-branes in the other two cases)
leads to anomaly-free results.  The case of  five-branes or small $SO(32)$
instantons was considered in (\cite{witsm}). When $n_B$ branes coincide  at
the same point (and away from singularities) a non-perturbative gauge  group
$Sp(n_B)$ is expected to appear, along with hypermultiplets transforming in
the fundamental, antisymmetric and singlet representations. We also assign
  these hypermultiplets into representations of the perturbative group. Thus,
the massless matter content, transforming under the  full $U(m)\times
SO(32-2m)\times Sp(n_B)$ group is
\beqa
& {} & \frac12({\bf m},{\bf 1}, {\bf 2n_B})+ \frac12({\bf \overline{m}},{\bf
1}, {\bf 2n_B})+  \frac12({\bf 1},{\bf 32-2m},{\bf 2n_B}) \nonumber \\
& + & ({\bf 1},{\bf 1},{\bf \frac{2n_B(2n_B-1)}2 -1}) + ({\bf 1},{\bf
1},{\bf 1}) ]]
\label{bcomple}
\eeqa
It is straightforward to check that all non-Abelian gauge and gravitational
anomalies do cancel.  Thus, our construction provides a new class of
  consistent  non-perturbative orbifold heterotic vacua.

Notice that the models obtained require the addition of  $6s$, $s=4,3,2,1,0$,
five-branes.  They contribute one unit of magnetic charge each.
Thus, in order
to achieve overall vanishing magnetic charge, each of the fixed
points (which in these  particular models are identical) must carry magnetic
charge $q_f=-{\frac{n_B}{9}}$.

The $E_8\times E_8$ case is to some extent similar
but has some peculiarities. Consider the class of models
generated by gauge shifts of the form
\beq
\label{shiftsee}
V \! = \! \frac13( 1,\! \cdots \!, 1,0, \! \cdots \!, 0)\times
\frac13 (1,\! \cdots \!, 1,0,\! \cdots \!,0)
\eeq
with an even number $m_1$ ($m_2$) of $\frac13$ entries in the first (second)
$E_8$
and w $m= m_1+m_2\leq 8$.  Models with appropriate oscillator
moduli in the twisted sector have
$(m_1,m_2)= (0,0)$, $(2,0)$,$(4,0)$, $(2,2)$, $(2,4)$ and $(4,4)$.
Again, none of these models (except for $(m_1,m_2)=(4,4)$)
fulfill the perturbative modular invariance constraints and
are, therefore, anomalous. However, unlike the $SO(32)$ case, they
{\it  do not present
non-Abelian gauge anomalies}. We can check that they miss an equivalent
of $3(8-m)\times 30$
hypermultiplets in order to cancel gravitational anomalies.
But this is precisely the contribution corresponding to
$3(8-m)$ M-theory five branes, each one carrying a tensor multiplet
and a gauge singlet hypermultiplets. Therefore,
these missing modes match the non-perturbative spectrum
corresponding to setting this same number of instantons to zero size
in $E_8\times E_8$.
This is a nice check of  our  procedure. Simple addition of a shift in
the vacuum energy automatically takes into account the difference between
the $SO(32)$ and $E_8\times E_8$ heterotic strings, yielding no gauge
anomalies in the second case.
The $Z_3$ models under consideration  are
orbifold realizations of the $E_8\times E_8$ vacua in the presence of
wandering branes considered in refs.\cite{dmw,sw6d,mv1}.

An interesting question is  whether
there is any shift $V$ in $E_8\times E_8$ (or $Spin(32)/Z_2$)
which admits both spectra, with and without
five-branes.  Such a situation, could  indicate possible
 transitions between perturbative and non-perturbative vacua which
proceed through the emission of five-branes to the  bulk.
Indeed,
there is a unique case corresponding to the `standard embedding',
$V=\frac 13(1,1,0,\cdots,0)\times (0, \cdots,0)$
($V=\frac13(1,1,0,\cdots,0)$ for $Spin(32)/Z_2$)  in which
there are both a model without five-branes and a model with 18 five-branes.
Both models have identical untwisted perturbative spectrum but differ
in that
the twisted spectrum  of the perturbative model has  extra hypermultiplets
with respect to the non perturbative one.

In the $E_8\times E_8$ case they transform as
$({\bf 56},{\bf 1})+7({\bf  1},{\bf 1})$ while they organize as
$({\bf 2},{\bf 28}) +4({\bf 1},{\bf 1})$  under $SO(28)\times U(2)$ for
$Spin(32)/Z_2$.
 The corresponding non-perturbative model contains just  three  singlets per
fixed point in both cases.  In the non-perturbative model
the fixed points have magnetic charge $Q_f=-2$.
This suggests that there can be transitions by which, around a fixed point
in  the perturbative model,  these hypermultiplets go into 2 five branes
producing the non perturbative model. The magnetic charge is conserved
during the process since each fixed point has charge $Q_f=-2$ and each of
the five-branes has charge +1.

In the $Spin(32)/Z_2$ case these transitions can be interpreted as an
unhiggsing process where the rank is increased by two units, namely
$(2,28) +4(1,1) \rightarrow Sp(2) +matter$. If this transition occurs at
each of the 9 fixed points and  all the branes are at the same (non singular
point) an $Sp(18)$ maximum enhanced group is obtained, with the matter
content specify in ~(\ref{bcomple}). A similar enhancing is expected to
occur in $D=4$.

  $E_8\times E_8$ case is  different. In the transition
$\ \ ({\bf 56},{\bf 1})+4({\bf 1},{\bf 1}) \rightarrow 2( {\bf 1}+tensor)$
there is no enhancing at all and a  complete charged hypermultiplet
disappears into the bulk. In terms of $M-theory$ branes this
corresponds to an $E_8$ instanton, living on one of the ``end of the world''
nine branes, becoming pointlike and going into the bulk as a five M-brane
\cite{sw6d,gh}.
Interestingly enough, if an equivalent transition was possible  in $D=4$,
for $N=1$ it would imply a change in the number of generations. A chiral
$\bf 27$ generation (or $\bf \overline { 27}$) of $E_6$, contained in
the ${\bf 56}$ of $E_7$ would disappear from the spectrum
We will show an explicit realization below.

Transitions  can happen at each fixed point independently  so that
there should exist similar models  with any even number of five-branes in
between 2 and 18.  Thus, in this standard embedding models there is a
discrete degree of freedom which corresponds to having  pairs of zero size
instantons.

Here we have concentrated in an certain class of $Z_3$ models
with enough blowing up modes to  resolve the singular points completely.
A more general situation can be envisaged for cases where these modes are
lacking (  $V^2\ > \frac89$ above) and for other $Z_M$ orbifolds.
This is extensively discussed in \cite{afiuv}. Let us just recall
that generically non-smooth models have five branes trapped at these non
removable singularities. The dynamics associated to these stuck branes is
different from that of the smooth case.  The behaviour of such five-branes
for the $SO(32)$ heterotic string is better known. It can be extracted from
type I D-five-branes on ALE spaces  and F-theory analysis.
\cite{quivers,intri,bi,bi2,aspfz2}.  For a bigger enough number $l_c$ of
branes sitting at a singularity, further enhancings to unitary groups are
expected. For instance, at a $Z_3$ orbifold point $S_p(l)\times U(2l+m)$
is obtained ($l_c=\frac{8-m}2$).  Tensor multiplets associated to the
 missing blowing up modes do appear, somehow paralleling the $E_8\times E_8$
case with wandering branes.
 Moreover, transitions where some hypermultiplets go into tensors are also
suggested. For instance, when $m=0$ and $l=0$ in the $Z_3$
there is no enhancing at all and it is found that
${\bf {28}} + {\bf {1}} \longrightarrow {\rm tensor,}$
where   ${\bf {28}}$ is a hypermultiplet transforming under
a perturbative $U(8) (\times SO(16))$ group.
This parallels the above  $E_8\times E_8$ example.

 The idea  explored  in the $D=6$  case could be extended
to $D=4, N=1$.  One  would
construct heterotic orbifold vacua
with perturbative and non-perturbative sectors
in which the perturbative (but non-modular invariant)
sector  could  be  understood  in terms of  simple standard orbifold
techniques.
We should  also add a non-perturbative piece, but we face the
problem that non-perturbative phenomena in $N=1$, $D=4$
theories are poorly understood at the moment.
However, we can concentrate \cite{afiuv} on
certain  restricted classes of $D=4$ orbifolds
in which much of the structure is expected to be inherited from $D=6$.
In particular, one can consider  $Z_N\times Z_M$ orbifolds  in $D=4$ with
unbroken $N=1$ supersymmetry.  Such type of orbifolds have two general
classes of twisted sectors, those that leave a 2-torus fixed and those
that only leave fixed points.  The first type of twisted sectors is
essentially 6-dimensional  in nature,  the twist by itself would lead to
an  $N=2$, $D=4$  theory which would correspond to $N=1$, $D=6$
upon decompactification of the fixed torus.  For this type of twisted sectors
we can use our knowledge of  non-perturbative $D=6$, $N=1$
dynamics.  Twisted
sectors of the second type are purely 4-dimensional in nature and we would
need extra  information about  4-dimensional non-perturbative dynamics.  To
circumvent this lack of knowledge, one can  restrict to a particular class
of $Z_N\times Z_M$  orbifolds  with gauge embeddings  such that these purely
4-dimensional twisted sectors are either absent or else are not expected to
modify   the structure of the model substantially .

Let us present  a specific example \cite{afiuv} based on $E_8\times E_8$.
Consider the  $Z_3\times Z_3$ orbifold on $E_8\times E_8$ with gauge shifts
\beqa
A & = & \frac13(1,1,0, \cdots ,0)\times (0, \cdots, 0) \nonumber \\
B & = & \frac13(0,1,1,0, \cdots, 0)\times (0, \cdots ,0)
\label{abex2}
\eeqa
This leads to a perfectly modular invariant orbifold with gauge group
$E_6\times U(1)^2\times E_8$.
However, we are going to consider the particular version of this orbifold
with discrete torsion first considered in Ref.[19].  This model
has the special property that all particles in the $(A+B)$  twisted sector,
 are projected out.  In this way we get rid of the sector which is purely
4-dimensional. The model has now three ${\bf 27}$'s in the untwisted sector
and nine ${\bf \overline {27}}$'s in each of the sectors $A,B$ and $A-B$.
Hence, altogether the model has twenty four net antigenerations. We can now
consider a  non-perturbative orbifold in which the $D=6$ subsectors $A,B$
and $A-B$ have a left-handed vacuum energy shifted by $\frac13$.  This
corresponds to a non-perturbative $D=6$ vacuum with just singlets in the
twisted sectors and eighteen five-branes (leading to tensor multiplets) in
each of the three twisted sectors.

Therefore, a transition from perturbative to the non perturbative one implies
\begin{equation}
3({\bf 27})+ 27( {\bf \overline {27}}) \rightarrow 3 ({\bf 27})
\end{equation}

 The twenty seven antigenerations of
the twisted sectors disappear into the bulk and we are
only left  with three $E_6$ generations coming from the untwisted sector,
plus singlets.
The $U(1)$'s will now be anomalous but there will be extra
chiral singlets, coming from the tensors, with non-universal couplings to
the gauge fields which will lead to a generalized version of the GS
mechanism in $D=4$.

Other examples undergoing chirality changes may be considered.
A similar situation is found in $Spin(32)/Z_2$ when there are branes
stuck at a fixed point (\cite{afiuv}). A similar $Z_3 \times Z_3$ orbifold
projection applied to the $U(8) \times SO(16)$ model mentioned above
 leads for instance to an $SU(6)\times SU(2)$ non
abelian gauge group where a transition
\begin{equation}
 ({\bf \overline {15}},{\bf 1})+({\bf {6}},{\bf 2})
\rightarrow singlets
\end{equation}
occurs. Notice that, an {\it anomaly free} representation disappears from
the spectrum.

\section{Comments and outlook}

Since duality connects heterotic models with models derived from
other formulations, heterotic string theory has lost the privilege of
being the
preferred theory for establishing links with phenomenological world.
In fact,  most of the models we have constructed have candidate duals
(this is a further check of our proposal) obtained
from $F-theory$, $ M-theory$ and type I string formulations\cite{afiuv}.

Nevertheless, it seems to us that, the  heterotic orbifolds constructions
we are proposing are particularly attractive for their simplicity, even
if other construction, like $F-theory$, could be more powerful.
This construction can be summarized in three steps: i. Find possible
shifts (in general automorphisms) on the gauge lattice, both satisfying the
constraint ~(\ref{dos}) or not . ii. Add an energy shift $E_B$ to ensure
level matching. Then, compute the perturbative spectrum by using familiar
orbifold techniques.  iii. Add non perturbative physics information. In $
D=6$ this  corresponds to branes wandering  in the bulk or  trapped at fixed
points, their world volume fills space time .  Even in $D=6$, this
last step is only partially known. As we stressed, only for a large enough
number of branes $l\ge l_c$  on a fixed point and for $Spin(32)/{Z_2}$
lattice, small instanton information is available.  This information is not
yet available for  $E_8\times E_8$.  Furthermore,  except for some cases,
non-perturbative spectrum in not known in either of both lattices,
when the number of small instantons on the singularity is smaller
than the critical value.  The situation for $D=4$, $N=1$ vacua is even more
uncertain. Some insight can be obtained from recent type IIB orientifold
constructions\cite{ori4,afiv}. Even though, we have seen that relevant non
perturbative information can be derived, in certain cases, from $D=6$
physics.   The examples exhibiting chirality
changing transitions are particularly interesting. They correspond to
$D=6$ transitions in which one tensor multiplet transmutes into  twenty nine
charged hypermultiplets.  These transitions where also studied in
Ref.[23] in another context.  These examples show that the number
of chiral generations is not invariant under non-perturbative effects.
Something inconceivable in perturbative field theory and  also in
perturbative string theory where the net number of generations is a
topological number associated to a given compactified internal manifold.
Vacua with different number of generations can be  connected. Even if the
processes involve strong coupling dynamics, quite presumably part or all the
connected four dimensional models can be realized perturbatively in some
region of moduli space thus, effectively reducing the excessively
huge vacuum degeneracy.  In the explicit examples we have sketched above,
these transitions may occur at each fixed point independently. If these are
achieved at all nine $Z_3$ fixed points we end up with three generations and
this number, associated to the untwisted sector of the orbifold,   cannot be
 reduced further. In perturbative string theory some effort has been
 dedicated to find appropriate compactifications leading to a small three,
 may be four (non vanishing) number of  generations, hoping that non
perturbative physics would privilege these realizations over infinitely many
others.  These transitions, indicate that, at least in some cases, the
 strongly coupled dynamics leading to models with few generations is
		  available.  Of course, other new phenomenological questions
should be taken up now.  For instance,  since in other
 compactifications two or zero net generations are obtainable,  which would
be the preferred number?  Other new, non perturbative, fact is that the gauge
group may be significantly enhanced. This enhancement may be amazingly huge,
and discouraging for predictivity,  as it was found in some very singular
F-theory compactifications\cite{rank}.  The situation is much more bounded
in the models we have discussed above.  In particular non perturbative
enhanced groups which contain factors  like $SU(n)\times SU(n) $ with matter
in $({\bf n},{\bf \overline {n}})$  are frequently found and they appear as
specially apt  for obtaining Grand Unified like models with adjoint
representations (needed for GUT symmetry breaking). These could be achieved
by giving adequate vev's to one of the above representations to obtain the
diagonal $SU(n)$ group.  Interestingly enough in [21] we find
that such kind of breaking can be achieved in a type IIB dual orientifold
construction through a consistent inclusion of continuous Wilson lines. For
instance, a $U(4)^3 (\times U(4)) $ gets broken to a diagonal
$U(4)_{diag} (\times U(4))$, in a $Z_3$ orientifold example,
through this mechanism (see also [22]).  Let us remark that, unlike
string GUTs constructions considered
before (see for instance [25] and references therein), the unified group
 here would be non-perturbative, from the heterotic point of view.  Also
notice that it is not clear how, non perturbative, fermionic generations
would arise.  In particular, spinorial representations of $SO(2n)$ (and so
 for instance a generation ${\bf 16}$ of $SO(10)$ containing a ${\bf 5} +
{\bf \overline {10}}$ of $SU(5)$ ) do not appear in the spectra associated
to the presence of pointlike instantons, up to the present partial knowledge
of the subject.  Again, new phenomenological questions must be faced, not
envisaged from perturbative heterotic string formulations.  We  conclude by
recalling that extensions of our symmetric orbifold construction might be
considered to asymmetric (non modular invariant) orbifolds or to arbitrary
conformal field theories.

My warm thanks to  A. Font, L. Iba\~nez, A. M. Uranga
and G. Violero, with whom this work was done, for a specially enjoyable
collaboration.  
\end{document}